\documentclass[aps,prb,twocolumn,amsmath,amssymb,showpacs]{revtex4}

\usepackage{graphicx}
\usepackage{dcolumn}
\usepackage{bm}

\begin{document}

\title{Multipole ordering in $f$-electron systems
on the basis of a $j$-$j$ coupling scheme}

\author{Katsunori Kubo}
\author{Takashi Hotta}

\affiliation{
Advanced Science Research Center, Japan Atomic Energy Research Institute,
Tokai, Ibaraki 319-1195, Japan}

\received{23 May 2005}
\published{3 October 2005}

\begin{abstract}
We investigate microscopic aspects of multipole ordering
in $f$-electron systems with emphasis on the effect of lattice structure.
For the purpose, first we construct $f$-electron models
on three kinds of lattices, simple cubic (sc), bcc, and fcc,
by including $f$-electron hopping through $(ff\sigma)$ bonding
in a tight-binding approximation
on the basis of a $j$-$j$ coupling scheme.
Then, an effective model is derived in the strong-coupling limit
for each lattice structure with the use of second-order perturbation
theory with respect to $(ff\sigma)$.
By applying mean-field theory to such effective models,
we find different types of multipole ordered state
depending on the lattice structure.
For the sc lattice, a $\Gamma_{3g}$ antiferro-quadrupole transition
occurs at a finite temperature and as further lowering temperature,
we find another transition to a ferromagnetic state.
For the bcc lattice, a $\Gamma_{2u}$ antiferro-octupole ordering
occurs first, and then, a ferromagnetic phase transition follows it.
Finally, for the fcc lattice, we find a single phase transition
to the longitudinal triple-$\bm{q}$ $\Gamma_{5u}$ octupole ordering.
\end{abstract}

\pacs{75.30.Et, 71.10.Fd, 75.40.Cx}


\maketitle

\section{Introduction}

It is one of currently important issues in the research field of
condensed matter physics to unveil exotic magnetic properties of
strongly correlated electron materials
with active orbital degree of freedom.
Among those materials, in $d$-electron systems such as
transition metal oxides, origin of complex magnetic structure
has been vigorously discussed based on the concept of
orbital ordering.\cite{Imada,Tokura,Dagotto,Hotta0}
Also in $f$-electron materials including rare-earth and
actinide elements, various kinds of magnetic and orbital ordering
have been found.\cite{Santini2,orbital2001}
It is now widely recognized that orbital degree of freedom
plays a crucial role for the emergence of novel magnetism
in $d$- and $f$-electron systems.

Here we should note that in $f$-electron systems,
spin and orbital are not independent degrees of freedom,
since they are tightly coupled with each other
due to the strong spin-orbit interaction.
Then, in order to describe such a complicated spin-orbital coupled system,
we usually represent the $f$-electron state in terms of
``multipole'' degree of freedom,
rather than using spin and orbital degrees of freedom
as in $d$-electron systems.
Among multipole moments, there have been intensive and extensive studies
on dipole and/or quadrupole ordering in $f$-electron systems.
In usual cases, magnetic ordering indicates dipole one,
which can be detected by neutron diffraction experiments.
Ordinary orbital ordering means quadrupole one,
which can also be detected experimentally,
since it induces lattice distortions
due to the spatial anisotropy in charge distribution.

In addition to dipole and quadrupole ordering,
in recent years, possibility of higher-order multipole ordering,
i.e., magnetic octupole ordering, has been also discussed for
Ce$_x$La$_{1-x}$B$_6$
\cite{Kuramoto,Kusunose,Sakakibara,Kubo,Kobayashi,Iwasa,Suzuki}
and NpO$_2$,
\cite{Santini,Paixao,Lovesey,Kiss,Tokunaga,Sakai2,Kubo:NpO2}
to reconcile experimental observations which seem to contradict
one another at first glance.
Very recently, a possibility of octupole ordering has been
proposed also for SmRu$_4$P$_{12}$.~\cite{Yoshizawa,Hachitani}
It is noted that in these materials, crystalline electric field (CEF)
ground states are $\Gamma_8$ quartets with large degeneracy
even under a CEF potential.~\cite{Zirngiebl,Fournier,Matsuhira}
In the $\Gamma_8$ ground-state multiplet, octupoles exist as
independent moments besides dipole and quadrupole moments.\cite{Shiina}
Then, phenomenological theories have been developed under the assumption
that octupole ordering occurs.
Note that direct detection of octupole ordering is very difficult,
since the octupole moment directly couples to neither a magnetic
field nor lattice distortions.
However, those phenomenological theories have been successful
in explaining several experimental facts consistently, e.g.,
induced quadrupole moments in octupole ordered states
in Ce$_x$La$_{1-x}$B$_6$~\cite{Kusunose,Kubo}
and NpO$_2$.\cite{Paixao,Kiss}

As mentioned above, thus far, the study on multipole ordering
in $f$-electron systems has been almost limited
in the phenomenological level, mainly due to the complexity
in the treatment of multipole degree of freedom.
It might be possible to consider a Heisenberg-like model for
multipole moments, but the interactions among multipole moments
were determined just phenomenologically.
It is highly required to proceed to microscopic theory,
in order to understand the origin of multipole ordering
in $f$-electron systems.
However, it is very hard and practically impossible to study
multipole ordering in the model retaining all the $f$-electron states.
Then, it is necessary to consider a tractable model which keeps
correct $f$-electron symmetry.

One way for such model construction is to use an $LS$ coupling scheme.
For instance, the Ruderman-Kittel-Kasuya-Yosida (RKKY) interactions
were estimated in DyZn,~\cite{Schmitt} and
in CeB$_6$ and CeB$_2$C$_2$~\cite{Sakurai}
from microscopic models using the $LS$ coupling scheme.
However, the method based on the $LS$ coupling scheme is
complicated and seems still hard to be extended.
One reason for the difficulty is that we cannot apply
standard quantum-field theoretical techniques
in the $LS$ coupling scheme, since Wick's theorem does not hold.
From this viewpoint, it is recommended to use a $j$-$j$ coupling
scheme.\cite{Hotta}
Since individual $f$-electron states are first defined, we can include
many-body effects in systematic ways using theoretical techniques
developed for the research of $d$-electron
systems.~\cite{Kubo:NpO2,Hotta,Hotta2,Hotta3a,Hotta3}

In this paper, in order to investigate how multipole ordering appears
in $f$-electron systems from a microscopic viewpoint,
we exploit the $j$-$j$ coupling scheme.
We construct tight-binding models on three kinds of
lattices, simple cubic (sc), bcc, and fcc,
by including Coulomb interactions among $\Gamma_8$ states.
In order to discuss multipole ordering in these models,
we derive an effective multipole interaction model
in the strong-coupling limit for each lattice structure
by using the second-order perturbation theory
with respect to $f$-$f$ hopping integrals,
as to estimate the superexchange interaction in $d$-electron systems.
Then, within a mean-field approximation,
we clarify what kind of multipole ordering occurs
in the effective model:
For the sc lattice, a $\Gamma_{3g}$ antiferro-quadrupole transition
occurs, while for the bcc lattice, $\Gamma_{2u}$ antiferro-octupole
ordering appears.
For the fcc lattice with geometrical frustration, we find
longitudinal triple-$\bm{q}$ $\Gamma_{5u}$ octupole ordering.

The organization of this paper is as follows.
In Sec.~II, we introduce a tight-binding model
based on the $j$-$j$ coupling scheme
including only the $\Gamma_8$ states.
In Sec.~III, we describe the general prescription to
derive an effective Hamiltonian from the $\Gamma_8$ model.
In Sec.~IV, we show the mean-field results of the effective models
on sc, bcc, and fcc lattices.
Finally, in Sec.~V, the paper is summarized.

\section{Hamiltonian}
\label{sec:model}

When we study theoretically the $f$-electron properties,
the $LS$ coupling scheme has been frequently used to include
the effect of Coulomb interactions, spin-orbit coupling,
and CEF potential.
However, as mentioned above, it is not possible to apply
standard quantum-field theoretical technique
in the $LS$ coupling scheme, since Wick's theorem does not hold.
In order to overcome such a difficulty, it has been proposed to
construct a microscopic model for $f$-electron systems
by exploiting the $j$-$j$ coupling scheme,\cite{Hotta}
where we include first the spin-orbit coupling so as
to define the state labeled by the total angular momentum $j$.
For $f$ orbitals with angular momentum $\ell$=3,
we immediately obtain an octet with $j$=7/2(=3+1/2) and
a sextet with $j$=5/2(=3$-$1/2),
which are well separated by the spin-orbit interaction.
Since the spin-orbital coupling is, at least, in the order
of 0.1 eV for $f$ electrons, it is enough to take into account
the $j$=5/2 sextet, when we investigate low-temperature properties
of $f$-electron compounds in the $j$-$j$ coupling scheme.

In order to construct the many-body state,
we accommodate $f$ electrons in the $j$=5/2 sextet
by following the Hund's rule interactions and CEF potential,
as we have done for $d$-electron systems.
It has been found that the many-electron state obtained
in the $j$-$j$ coupling scheme is continuously changed
to the corresponding state in the $LS$ coupling scheme,
as long as those states in both schemes
belong to the same symmetry group.\cite{Hotta3}
Namely, if we based on the spirit of adiabatic continuation,
there is no serious difference between the states of the $LS$
and $j$-$j$ coupling schemes.
Depending on the problem, we can use one of the
schemes for $f$-electron systems.
For instance, if we attempt to explain phenomenologically
the experimental results of $f$-electron insulators,
it is highly recommended to use the $LS$ coupling scheme.
On the other hand,
the $j$-$j$ coupling scheme is rather appropriate to develop
a microscopic theory for novel magnetism and
unconventional superconductivity of $f$-electron systems.
In the present paper, our purpose is to construct a microscopic
theory for multipole ordering from the viewpoint of spin-orbital
complex.
Thus, we exploit the $j$-$j$ coupling scheme throughout this paper.

As described above, we consider only the states with $j$=5/2.
The $j$=5/2 states are further split into $\Gamma_7$ doublet and
$\Gamma_8$ quartet due to a cubic CEF.
In order to consider multipole phenomena such as octupole ordering
in $f$-electron systems from a microscopic viewpoint,
in this paper we consider only $\Gamma_8$ states by assuming large CEF
splitting energy between $\Gamma_7$ and $\Gamma_8$ levels.
This simplification is motivated by the fact that
the possibility of exotic octupole ordering has been actively
discussed in Ce$_x$La$_{1-x}$B$_6$ and NpO$_2$
with $\Gamma_8$ ground state.

Here readers may be doubtful of the reality of our assumption,
since the Coulomb interaction among $f$ electrons is naively thought
to be larger than the CEF level splitting in any case.
However, it should be noted that we are now considering
the $f$-electron state in the $j$-$j$ coupling scheme,
not in the original $f$-electron state with angular momentum $\ell$=3.
As pointed out in Ref.~\onlinecite{Hotta}, the Hund's rule interaction
in the $j$-$j$ coupling scheme is effectively reduced to be 1/49
of the original Hund's rule coupling.
Namely, even if the original Hund's rule coupling among $f$ electrons
is 1~eV, it is reduced to 200~K in the $j$-$j$ coupling scheme.
We note that the CEF level splitting in actinide dioxides
is considered to be larger than 1000~K.~\cite{Kubo:NpO2,Kubo:fp}
We also recall that the CEF level splitting in CeB$_6$
is as large as 500~K.~\cite{Zirngiebl}
Thus, we safely conclude that our present assumption is correctly
related to the realistic situation.
Of course, in order to achieve quantitative agreement with experimental
results, it is necessary to include also $\Gamma_7$ level,
since the magnitude of the CEF splitting is always finite,
even if it is large compared with the effective Hund's rule interaction.
However, we strongly believe that it is possible to grasp
microscopic origin of multipole ordering in $f$-electron systems
on the basis of the $\Gamma_8$ model,
since this model is considered to be connected adiabatically
from the realistic situation.
We postpone further effort to develop more general theory
to include all the $j$=5/2 sextet in future.

\begin{figure}
  \includegraphics[width=1\linewidth]{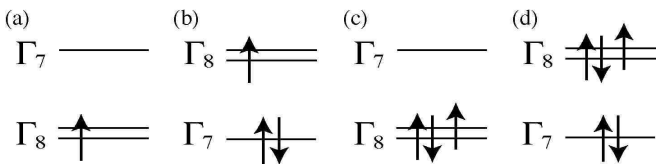}
  \caption{\label{figure:level_schemes}
    Electron configurations in the $j$-$j$ coupling scheme
    for $\Gamma_8$ CEF ground states.
    (a) One electron in the $\Gamma_8$ for $n$=1.
    (b) One electron in the $\Gamma_8$ for $n$=3.
    (c) One hole in the $\Gamma_8$ for $n$=3.
    (d) One hole in the $\Gamma_8$ for $n$=5.
  }
\end{figure}

Concerning the $f$-electron number, in this paper we treat only
the case with one $f$ electron in the $\Gamma_8$ multiplet per site.
However, this restriction does $not$ simply indicate that
we consider only the Ce-based compound.
In the $j$-$j$ coupling scheme, in order to consider $f^n$-electron
systems, where $n$ indicates local $f$ electron number per site,
we accommodate $f$ electrons in the one-electron CEF levels
due to the balance between Coulomb interactions and CEF level
splitting energy, just as in the case of $d$-electron systems.
Thus, the situation with one $f$ electron in the $\Gamma_8$ multiplet
per site expresses both cases with $n$=1 in the $\Gamma_8$-$\Gamma_7$
[Fig.~\ref{figure:level_schemes}(a)] and
$n$=3 in the $\Gamma_7$-$\Gamma_8$
[Fig.~\ref{figure:level_schemes}(b)] systems,
where $\Gamma_x$-$\Gamma_y$ symbolically denotes the situation
with $\Gamma_x$ ground and $\Gamma_y$ excited states.
Furthermore, we should note that due to the electron-hole symmetry
in the $\Gamma_8$ subspace, the effective model with one $f$ electron
in the $\Gamma_8$ state is the same for that in the case with
three electrons in the $\Gamma_8$ multiplet.
Namely, the present model also indicates both cases
with $n$=3 in the $\Gamma_8$-$\Gamma_7$
[Fig.~\ref{figure:level_schemes}(c)]
and $n$=5 in the $\Gamma_7$-$\Gamma_8$
[Fig.~\ref{figure:level_schemes}(d)] systems.

Before proceeding to the exhibition of the Hamiltonian,
it is necessary to define $f$-electron operators
in $\Gamma_8$ states.
Since the $\Gamma_8$ quartet consists of two Kramers doublets,
we introduce orbital index $\tau$ (=$\alpha$ and $\beta$)
to distinguish the two Kramers doublets,
while spin index $\sigma$ (=$\uparrow$ and $\downarrow$)
is defined to distinguish the two states in each Kramers doublet.
In the second-quantized form,
annihilation operators for $\Gamma_8$ electrons are defined as
\begin{subequations}
\begin{align}
  f_{\mathbf{r} \alpha \uparrow}
  &= \sqrt{5/6} a_{\mathbf{r} 5/2}+\sqrt{1/6} a_{\mathbf{r} -3/2},\\
  f_{\mathbf{r} \alpha \downarrow}
  &= \sqrt{5/6} a_{\mathbf{r} -5/2}+\sqrt{1/6} a_{\mathbf{r} 3/2},
\end{align}
\end{subequations}
for $\alpha$-orbital electrons, and
\begin{subequations}
\begin{align}
  f_{\mathbf{r} \beta \uparrow} &= a_{\mathbf{r}  1/2},\\
  f_{\mathbf{r} \beta \downarrow} &= a_{\mathbf{r} -1/2},
\end{align}
\end{subequations}
for $\beta$-orbital electrons, where $a_{\mathbf{r} j_z}$ is
the annihilation operator for an electron with the $z$-component
$j_z$ of the total angular momentum at site $\mathbf{r}$.

Now we show the Hamiltonian of $\Gamma_8$ electrons.
For the purpose to consider the effective model later,
it is convenient to express the Hamiltonian in the form of
\begin{equation}
  \label{eq:H}
  \mathcal{H}=\mathcal{H}_{\text{kin}}+\mathcal{H}_{\text{loc}},
\end{equation}
where $\mathcal{H}_{\rm kin}$ denotes the kinetic term of $f$ electrons
and $\mathcal{H}_{\text{loc}}$ indicates the local interaction
part for $\Gamma_8$ electrons.
In this paper, the kinetic term of $\Gamma_8$ electrons is given
by exploiting the tight-binding approximation.
Then, $\mathcal{H}_{\text{kin}}$ is expressed as
\begin{equation}
  \mathcal{H}_{\text{kin}}
  =\sum_{\mathbf{r},\bm{\mu},\tau,\sigma,\tau^{\prime},\sigma^{\prime}}
  t^{\bm{\mu}}_{\tau \sigma; \tau^{\prime} \sigma^{\prime}}
  f^{\dagger}_{\mathbf{r} \tau \sigma}
  f_{\mathbf{r}+\bm{\mu} \tau^{\prime} \sigma^{\prime}},
\end{equation}
where $\bm{\mu}$ is a vector connecting nearest-neighbor sites
and $t^{\bm{\mu}}_{\tau \sigma; \tau^{\prime} \sigma^{\prime}}$
is the hopping integral of an electron with
$(\tau^{\prime}, \sigma^{\prime})$ at site $\mathbf{r}$+$\bm{\mu}$
to the $(\tau, \sigma)$ state at $\mathbf{r}$.
We note that the hopping integral
$t^{\bm{\mu}}_{\tau \sigma; \tau^{\prime} \sigma^{\prime}}$
depends on orbital, spin, and direction $\bm{\mu}$,
due to $f$-electron symmetry.
Then, the form of the hopping integral is characteristic of
lattice structure.
The explicit form of the hopping matrix will be shown later
for each lattice structure.
Note also the relation
$t^{-\bm{\mu}}_{\tau \sigma; \tau^{\prime} \sigma^{\prime}}$
=$t^{\bm{\mu}}_{\tau \sigma; \tau^{\prime} \sigma^{\prime}}$.

As for the local $f$-electron term $\cal{H}_{\rm loc}$,
since we assume the large CEF splitting energy
between $\Gamma_7$ and $\Gamma_8$ levels,
it is enough to consider the Coulomb interaction terms
among $\Gamma_8$ electrons.
As easily understood from the introduction of
`spin' and `orbital' in the $j$-$j$ coupling scheme,
the local $f$-electron term in the $\Gamma_8$ quartet
becomes the same as that of the two-orbital systems
for $d$ electrons.
In fact, after lengthy algebraic calculations for Racah parameters
in the $j$-$j$ coupling scheme,\cite{Hotta}
$\cal{H}_{\rm loc}$ is given as
\begin{equation}
  \begin{split}
    \mathcal{H}_{\text{loc}}
    &=
    U \sum_{\mathbf{r} \tau}
    n_{\mathbf{r} \tau \uparrow} n_{\mathbf{r} \tau \downarrow}
    +U^{\prime} \sum_{\mathbf{r}}
    n_{\mathbf{r} \alpha} n_{\mathbf{r} \beta}
    \\
    &+ J \sum_{\mathbf{r},\sigma,\sigma^{\prime}}
    f^{\dagger}_{\mathbf{r} \alpha \sigma}
    f^{\dagger}_{\mathbf{r} \beta \sigma^{\prime}}
    f_{\mathbf{r} \alpha \sigma^{\prime}}
    f_{\mathbf{r} \beta \sigma}
    \\
    &+ J^{\prime}\sum_{\mathbf{r},\tau \ne \tau^{\prime}}
    f^{\dagger}_{\mathbf{r} \tau \uparrow}
    f^{\dagger}_{\mathbf{r} \tau \downarrow}
    f_{\mathbf{r} \tau^{\prime} \downarrow}
    f_{\mathbf{r} \tau^{\prime} \uparrow},
  \end{split}
  \label{f_local}
\end{equation}
where $n_{\mathbf{r} \tau \sigma}$
=$f^{\dagger}_{\mathbf{r} \tau \sigma} f_{\mathbf{r} \tau \sigma}$
and
$n_{\mathbf{r} \tau}$=$\sum_{\sigma} n_{\mathbf{r} \tau \sigma}$.
The coupling constants $U$, $U^{\prime}$, $J$, and $J^{\prime}$
denote the intra-orbital Coulomb, inter-orbital Coulomb, exchange,
and pair-hopping interactions, respectively.
These are expressed in terms of Racah parameters, and we obtain
the relation $U$=$U'$+$J$+$J'$,  which can be understood from
the rotational invariance in orbital space.\cite{Hotta}
Note that for $d$-electron systems, one also has the relation
$J$=$J'$. When the electronic wave-function is real, this relation
is easily demonstrated from the definition of the Coulomb integral.
However, in the $j$-$j$ coupling scheme the wave-function is
complex, and $J$ is not equal to $J'$ in general.

\section{Effective Model}
\label{sec:effectiveH}

In this section, we describe a method to derive an effective Hamiltonian
by using the second-order perturbation theory with respect to
hopping integrals.
Here we emphasize that the procedure is essentially the same as
to estimate superexchange interactions for $d$-electron systems,
although the calculations are tedious due to the existence of
orbital degree of freedom.
After that, we will apply the standard mean-field theory to the
effective model to depict the phase diagram including
multipole ordered states.
We believe that it is meaningful to understand
the complicated $f$-electron multipole problem by using a simple
$d$-electron-like procedure and approximations,
both from conceptual and practical viewpoints.

When the Hamiltonian is written in the form of Eq.~\eqref{eq:H},
first we solve the local problem
\begin{equation}
  {\cal H}_{\rm loc}|\Phi^a_{n} \rangle = E_n|\Phi^a_{n} \rangle,
\end{equation}
where $E_n$ denotes the $n$-th eigenenergy and $|\Phi^a_{n} \rangle$
is the corresponding eigenstate with a label $a$ to distinguish
the degenerate states.
Since we accommodate one electron per site,
the ground state $|\Phi^a_{0} \rangle$ is expressed as
\begin{equation}
  |\Phi^a_{0} \rangle = \prod_{\mathbf{r},\tau,\sigma}
  f^{\dagger P_a(\mathbf{r},\tau,\sigma)}_{\mathbf{r} \tau \sigma}
  |0\rangle,
\end{equation}
where $|0 \rangle$ is the vacuum state of $f$ electrons,
$a$ denotes the electron configuration
in the ground state with one electron per site,
and $P_a(\mathbf{r},\tau,\sigma)$ takes 0 or 1
depending on the configuration $a$.

Here we consider the formal perturbation expansion in terms of
${\cal H}_{\rm kin}$ in order to construct the effective model.
Within the second order, ${\cal H}_{\rm eff}$ is generally written as
\begin{equation}
 \label{eq:Heff}
 {\cal H}_{\rm eff}
 = \sum_{a,b,u}\sum_{m \ne 0}
 |\Phi_0^a \rangle \langle \Phi_0^a| {\cal H}_{\rm kin}
 \frac{|\Phi^u_{m} \rangle \langle \Phi^u_{m}|}{E_0-E_m}
 {\cal H}_{\rm kin} |\Phi^b_0 \rangle \langle \Phi^b_0|,
\end{equation}
where $a$ and $b$ are labels to distinguish the ground states,
while $u$ is the label for the degenerate excited states.

Since we consider the situation with one electron per site,
the intermediate state due to one-electron hopping has
a vacant and a double occupied site.
The double occupied site has six possible states,
composed of $\Gamma_5$ triplet with energy $U'$$-$$J$,
$\Gamma_3$ doublet with energy $U'$+$J$(=$U$$-$$J'$),
and $\Gamma_1$ singlet with energy $U$+$J'$.
For the mathematical completion, it is necessary to include
all possible excited states in the intermediate process,
but the calculation becomes complicated.
Then, in this paper, in order to grasp the essential point
of the $\Gamma_8$ model by avoiding tedious calculations,
we include only the lowest-energy $\Gamma_5$ triplet
among the intermediate $f^2$ states.
This restriction to the intermediate states is validated,
when $J$ is much larger than the hopping energy of $f$ electron.
Since the $f$-electron hopping amplitude is considered to be
small compared with $J$, even if we also include the hybridization
with conduction electrons,
this approximation is acceptable in $f$-electron systems.

\begingroup
\squeezetable
\begin{table*}
  \caption{
  \label{table:multipole_operators}
  Multipole operators in the $\Gamma_8$ subspace.~\cite{Shiina}
  The site label $\mathbf{r}$ is suppressed in this Table for simplicity.
  }
  \begin{ruledtabular}
    \begin{tabular}{cccccccccccccccc}
      $\Gamma_{\gamma}$
      & $2u$   & $3gu$  & $3gv$
      & $4u1x$ & $4u1y$ & $4u1z$
      & $4u2x$ & $4u2y$ & $4u2z$
      & $5ux$  & $5uy$  & $5uz$
      & $5gx$  & $5gy$  & $5gz$ \\

      multipole operator $X^{\Gamma_{\gamma}}$
      & $T_{xyz}$   & $O^0_2$     & $O^2_2$
      & $J^{4u1}_x$ & $J^{4u1}_y$ & $J^{4u1}_z$
      & $J^{4u2}_x$ & $J^{4u2}_y$ & $J^{4u2}_z$
      & $T^{5u}_x$  & $T^{5u}_y$  & $T^{5u}_z$
      & $O_{yz}$    & $O_{zx}$    & $O_{xy}$ \\

      pseudospin representation
      & $\hat{\tau}^y$
      & $\hat{\tau}^z$
      & $\hat{\tau}^x$

      & $\hat{\sigma}^x$
      & $\hat{\sigma}^y$
      & $\hat{\sigma}^z$

      & $\hat{\eta}^+ \hat{\sigma}^x$
      & $\hat{\eta}^- \hat{\sigma}^y$
      & $\hat{\tau}^z \hat{\sigma}^z$

      & $\hat{\zeta}^+ \hat{\sigma}^x$
      & $\hat{\zeta}^- \hat{\sigma}^y$
      & $\hat{\tau}^x  \hat{\sigma}^z$

      & $\hat{\tau}^y \hat{\sigma}^x$
      & $\hat{\tau}^y \hat{\sigma}^y$
      & $\hat{\tau}^y \hat{\sigma}^z$ \\
    \end{tabular}
  \end{ruledtabular}
\end{table*}
\endgroup

Let us explain the prescription to derive the effective model
in the present case.
It is convenient to consider exchange processes of electrons
between two sites, $\mathbf{r}$ and $\mathbf{r}'$.
Since we consider the situation with one $f$ electron per site,
the initial state $|\mathbf{r} s_1,\mathbf{r}' s_2 \rangle$
is written as
\begin{equation}
  |\mathbf{r} s_1, \mathbf{r}' s_2 \rangle=
  f^{\dagger}_{\mathbf{r} s_1}
  f^{\dagger}_{\mathbf{r}' s_2}
  | 0 \rangle,
\end{equation}
where $s_1$ and $s_2$ symbolically denote spin and orbital states
for both electrons.
Then, we move one electron from the site $\mathbf{r}^{\prime}$ to
$\mathbf{r}$.
As mentioned above, the intermediate $f^2$ states at the site
$\mathbf{r}$ is restricted only as the lowest-energy $\Gamma_5$
triplet states.
Namely, the intermediated states should be expressed as
$|u, \mathbf{r} \rangle$ with the label $u$ to distinguish
the triplet states, given by
\begin{subequations}
\begin{align}
  |+1, \mathbf{r} \rangle &=
  f^{\dagger}_{\mathbf{r} \alpha \uparrow}
  f^{\dagger}_{\mathbf{r} \beta  \uparrow}
  |0 \rangle, \\
  |0, \mathbf{r} \rangle &=
   (f^{\dagger}_{\mathbf{r} \alpha \uparrow}
   f^{\dagger}_{\mathbf{r} \beta  \downarrow}
   +f^{\dagger}_{\mathbf{r} \alpha \downarrow}
   f^{\dagger}_{\mathbf{r} \beta  \uparrow})
  |0 \rangle /\sqrt{2}, \\
  |-1,\mathbf{r} \rangle &=
  f^{\dagger}_{\mathbf{r} \alpha \downarrow}
  f^{\dagger}_{\mathbf{r} \beta  \downarrow}
  |0 \rangle.
\end{align}
\end{subequations}

In order to obtain the effective model Eq.~(\ref{eq:Heff}),
it is enough to evaluate the inner product
\begin{equation}
  P_{u; s, s^{\prime}}
  =\langle \mathbf{r}, u 
  |\mathbf{r} s,\mathbf{r} s^{\prime} \rangle.
\end{equation}
This quantity is explicitly given by
$P_{+1; \alpha \uparrow \beta \uparrow}$=1,
$P_{ 0; \alpha \uparrow \beta \downarrow}$=$1/\sqrt{2}$,
and the other non-zero elements are given by
$P_{u;s^{\prime}s}$=$-P_{u;ss^{\prime}}$
and $P_{-u;\tau -\sigma \tau^{\prime} -\sigma^{\prime}}$
=$P_{u;\tau \sigma \tau^{\prime} \sigma^{\prime}}$.

Then, by including the processes in which an electron at $\mathbf{r}$
moves first, we obtain the effective Hamiltonian as
\begin{equation}
  \mathcal{H}_{\rm eff}
  =-\sum_{\langle \mathbf{r},\mathbf{r}^{\prime} \rangle}
  \sum_{s_1\text{--}s_4}
  I^{\mathbf{r}^{\prime}-\mathbf{r}}_{s_3, s_4; s_1, s_2}
  f^{\dagger}_{\mathbf{r} s_3}
  f_{\mathbf{r} s_1}
  f^{\dagger}_{\mathbf{r}^{\prime} s_4}
  f_{\mathbf{r}^{\prime} s_2},
  \label{two_site_interaction}
\end{equation}
where $\langle \mathbf{r},\mathbf{r}^{\prime} \rangle$
denotes the pair of nearest-neighbor sites
and the generalized exchange interaction $I$ is given by
\begin{equation}
  \begin{split}
  I^{\mathbf{r}^{\prime}-\mathbf{r}}_{s_3, s_4; s_1, s_2}
  &=
  \sum_{u,s,s^{\prime}}
      [(t^{\mathbf{r}^{\prime}-\mathbf{r}}_{s^{\prime}; s_4})^*
        P^*_{u; s_3, s^{\prime}}
        P_{u; s_1, s}
        t^{\mathbf{r}^{\prime}-\mathbf{r}}_{s; s_2}\\
        &+
        (t^{\mathbf{r}-\mathbf{r}^{\prime}}_{s^{\prime}; s_3})^*
        P^*_{u; s_4, s^{\prime}}
        P_{u; s_2, s}
        t^{\mathbf{r}-\mathbf{r}^{\prime}}_{s; s_1}]/(U^{\prime}-J).
  \end{split}
  \label{two_site_interaction_matrix}
\end{equation}

In order to investigate the multipole ordering, it is more convenient
to express the effective Hamiltonian Eq.~\eqref{two_site_interaction}
in terms of multipole operators.
For the purpose, we introduce some notations
to describe multipole operators as
\begin{subequations}
  \begin{align}
    \tilde{1}_{\tau \sigma; \tau^{\prime} \sigma^{\prime}}
    &\equiv \delta_{\tau \tau^{\prime}} \delta_{\sigma \sigma^{\prime}},\\
    \tilde{\bm{\tau}}_{\tau \sigma; \tau^{\prime} \sigma^{\prime}}
    &\equiv \bm{\sigma}_{\tau \tau^{\prime}} \delta_{\sigma \sigma^{\prime}},\\
    \tilde{\bm{\sigma}}_{\tau \sigma; \tau^{\prime} \sigma^{\prime}}
    &\equiv \delta_{\tau \tau^{\prime}} \bm{\sigma}_{\sigma \sigma^{\prime}},\\
    \tilde{\eta}^{\pm}
    &\equiv (\pm \sqrt{3}\tilde{\tau}^x-\tilde{\tau}^z)/2,\\
    \tilde{\xi}^{\pm}
    &\equiv -(\tilde{\tau}^x \pm \sqrt{3}\tilde{\tau}^z)/2,
  \end{align}
\end{subequations}
where $\bm{\sigma}$ are the Pauli matrices.
By using these notations,
we define one-particle operators at site $\mathbf{r}$ as
\begin{equation}
 \hat{A}_{\mathbf{r}} \equiv \sum_{\tau \tau^{\prime} \sigma \sigma^{\prime}}
 f^{\dagger}_{\mathbf{r} \tau \sigma}
 \tilde{A}_{\tau \sigma; \tau^{\prime} \sigma^{\prime}}
 f_{\mathbf{r} \tau^{\prime} \sigma^{\prime}},
\end{equation}
where $\tilde{A}$ is a $4 \times 4$ matrix.
The multipole operators in the $\Gamma_8$ subspace
are listed in Table~\ref{table:multipole_operators}.

With the use of above multipole operators,
the effective Hamiltonian is finally arranged in the form of
\begin{equation}
  \mathcal{H}_{\text{eff}}=
  \sum_{\mathbf{q}} (\mathcal{H}_{1 \mathbf{q}}
  +\mathcal{H}_{2 \mathbf{q}}
  +\mathcal{H}_{4u1 \mathbf{q}}+\mathcal{H}_{4u2 \mathbf{q}}),
  \label{eq:effectiveH}
\end{equation}
where $\mathbf{q}$ is the wave vector and
$\mathcal{H}_{1 \mathbf{q}}$ denotes quadrupole interactions.
$\mathcal{H}_{4un \mathbf{q}}$ ($n$=1 or 2)
denotes interactions between $\Gamma_{4un}$ moments and
ones between $\Gamma_{4un}$ and other octupole moments
with symmetry different from $\Gamma_{4u}$.
$\mathcal{H}_{2 \mathbf{q}}$ denotes other dipole
and octupole interactions.
In general, $\mathcal{H}_{2 \mathbf{q}}$ includes interactions
between $\Gamma_{4u1}$ and $\Gamma_{4u2}$ moments,
but we find that such interactions are not included
in the models with hopping integrals only through $(ff\sigma)$ bonding
on sc, bcc, and fcc lattices.
The explicit form of each multipole interaction
sensitively depends on the lattice structure,
as shown in the next section.

\section{Results}

Now we can calculate the effective interaction between two electrons
located along any direction $\mathbf{r}^{\prime}-\mathbf{r}$
by using Eq.~\eqref{two_site_interaction_matrix},
if the hopping integral along this direction is determined.
The hopping integrals of $f$ electrons
are evaluated by using the Slater-Koster table.~\cite{Takegahara}
In this section, we consider the nearest-neighbor hopping integrals
through $(ff\sigma)$ bonding for three lattice structures,
sc, bcc, and fcc.
Then, we present the effective Hamiltonian
and its ordered states for each lattice.
The structure of our effective model is consistent with
the general form of nearest-neighbor multipole interactions
on each lattice derived by Sakai~\textit{et al.}~\cite{Sakai}
We follow the notation in Ref.~\onlinecite{Sakai} for convenience.

\subsection{sc lattice}
\label{sec:sc}


The nearest-neighbor hopping integrals through $(ff\sigma)$ bonding
for the sc lattice are given by
\begin{subequations}
\begin{align}
  t^{(a,0,0)} &= [\tilde{1}-\tilde{\eta}^+]t_1,\\
  t^{(0,a,0)} &= [\tilde{1}-\tilde{\eta}^-]t_1,\\
  t^{(0,0,a)} &= [\tilde{1}-\tilde{\tau}^z]t_1,
\end{align}
\label{eqs:hopping:sc}
\end{subequations}
where $a$ is the lattice constant and $t_1$=$3(ff\sigma)/14$.


\begin{table}
  \caption{\label{table:sc_coupling_constants}
    Coupling constants in the effective model for the sc lattice.
    The energy unit is $(1/8)t_1^2/(U^{\prime}-J)$.
  }
  \begin{ruledtabular}
     \begin{tabular}{cccccccc}
       $a_1$ & $b_6$ & $b^{(1)}_1$ & $b^{(1)}_2$ & $b^{(1)}_3$ &
       $b^{(2)}_1$ & $b^{(2)}_2$ & $b^{(2)}_3$ \\
         12  &    3  &        $-4$ &        $-4$ &          0  &
                  4  &          1  & $-\sqrt{3}$
     \end{tabular}
  \end{ruledtabular}
\end{table}

For the sc lattice, the quadrupole interaction term
in Eq.~(\ref{eq:effectiveH}) is given by
\begin{equation}
    \mathcal{H}_{1 \mathbf{q}}
    =a_1(O^0_{2, -\mathbf{q}} O^0_{2, \mathbf{q}} C_z + \text{c.p.}),
    \label{eq:sc_effectiveH_even}
\end{equation}
where c.p. denotes cyclic permutations
and $C_{\nu}$=$\cos(q_{\nu} a)$ ($\nu$=$x$, $y$, or $z$).
The value of the coupling constant $a_1$
is given in Table~\ref{table:sc_coupling_constants}.

Note that $O^0_{2 \mathbf{q}}$ transforms to
$(\sqrt{3}O^2_{2 \mathbf{q}}-O^0_{2 \mathbf{q}})/2$
and $(-\sqrt{3}O^2_{2 \mathbf{q}}-O^0_{2 \mathbf{q}})/2$ under c.p.
$(x,y,z)\rightarrow(y,z,x)$ and $(x,y,z)\rightarrow(z,x,y)$, respectively.
The dipole and octupole interactions are given by
\begin{equation}
    \mathcal{H}_{2 \mathbf{q}}
    =b_6[T^{5u}_{z, -\mathbf{q}} T^{5u}_{z, \mathbf{q}} (C_x + C_y)
    +\text{c.p.}],
\end{equation}
and
\begin{equation}
  \begin{split}
    \mathcal{H}_{4u n \mathbf{q}}
    =b^{(n)}_1&[J^{4u n}_{z -\mathbf{q}} J^{4u n}_{z \mathbf{q}} C_z
      +\text{c.p.}] \\
    +b^{(n)}_2&[J^{4u n}_{z -\mathbf{q}} J^{4u n}_{z \mathbf{q}} (C_x + C_y)
      +\text{c.p.}] \\
    +b^{(n)}_3&[T^{5u}_{z -\mathbf{q}} J^{4u n}_{z \mathbf{q}} (C_x-C_y)
      +\text{c.p.})],
    \label{eq:sc_effectiveH_4u}
  \end{split}
\end{equation}
where values of the coupling constants $b_i$ and $b^{(n)}_i$
are shown in Table~\ref{table:sc_coupling_constants}.


Note that the form of the hopping integrals Eqs.~\eqref{eqs:hopping:sc}
are exactly the same as those for the $e_g$ orbitals
of $d$ electrons via $(dd\sigma)$ bonding.~\cite{Hotta,Anderson}
Thus, the effective Hamiltonian has the same form as in the $e_g$ model
considering only the lowest-energy intermediate states,~\cite{Kugel}
when we interpret that $\tau$ and $\sigma$ denote $e_g$ orbital
and real spin, respectively.
However, the physical meaning of the present model is
different from that of the $e_g$ model.
In particular, the effect of a magnetic field is essentially
different.
The dipole moment which couples to a magnetic field $\mathbf{H}$
is given by
$\mathbf{J}$=$(7/6)[\mathbf{J}^{4u1}$+$(4/7)\mathbf{J}^{4u2}]$
for the $\Gamma_8$ model, while for the $e_g$ model,
real spin $\bm{\sigma}$ of $d$ electrons is simply coupled
to a magnetic field.
In contrast to the $e_g$ model, a magnetic field resolves
the degeneracy in the $\tau$ space even within a mean-field theory
for the present model, as we will see later.


By applying mean-field theory to the effective model,
we find a $\Gamma_{3g}$ antiferro-quadrupole transition
at a temperature $T$=$T_{3g}$=$3a_1/k_{\text{B}}$.
As lowering temperature further,
we find a $\Gamma_{4u1}$ ferromagnetic transition.
This ferromagnetic transition can be regarded as a $\Gamma_{5u}$
antiferro-octupole transition,
since the $\Gamma_{4u1}$ ferromagnetic state with the $\Gamma_{3g}$
antiferro-quadrupole moment
is equivalent to the $\Gamma_{5u}$ antiferro-octupole ordered state
with the $\Gamma_{3g}$ antiferro-quadrupole moment.
The ground state energy is
$(-3/2)a_1-2b_6+b^{(1)}_1+2b^{(1)}_2$ per site.


\begin{figure}[t]
  \includegraphics[width=0.9\linewidth]{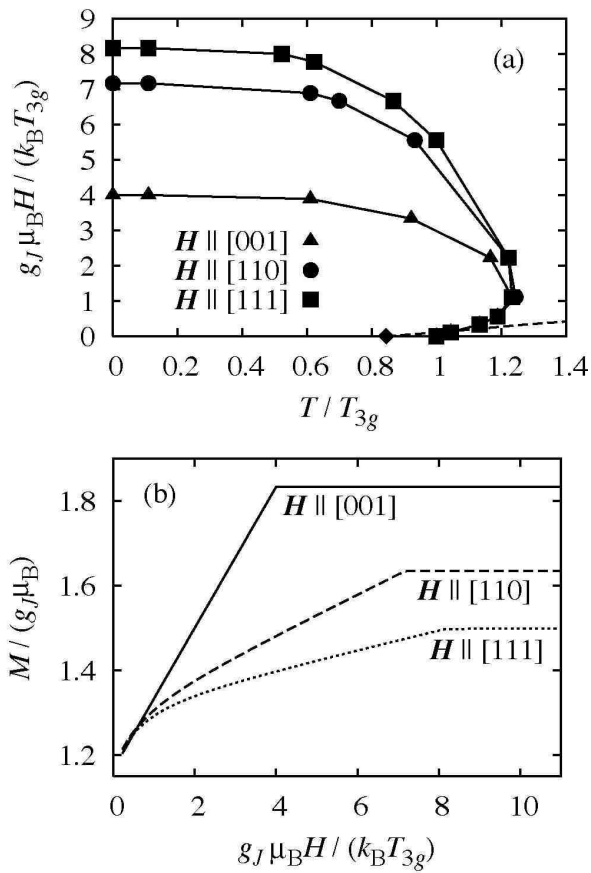}
  \caption{\label{figure:sc_PD_M}
    Phase diagram and magnetization for the sc lattice.
    The Land\'{e} $g$-factor is $g_J$=$6/7$.
    (a) $H$-$T$ phase diagram for three magnetic field directions.
    Solid symbols denote the $\Gamma_{3g}$ quadrupole transition.
    The diamond represents the ferromagnetic transition point.
    The dashed curve represents the crossover to the ferromagnetic state.
    The definition of the crossover is given in the main text.
    (b) Magnetization as a function of magnetic field.
  }
\end{figure}

In Fig.~\ref{figure:sc_PD_M}(a), we depict an $H$-$T$ phase diagram.
We note that the ferromagnetic transition at zero magnetic field
turns to be a crossover under the finite magnetic field.
The crossover is drawn by dashed curve,
determined by the peak position in the magnetic susceptibility.
Since it is found that the crossover curve is almost isotropic
in the region shown here,
we depict only the curve for $\mathbf{H} \parallel [001]$.
Note also that under a magnetic field, $\Gamma_{4u1}$ moments become finite,
and then, the $\Gamma_{5u}$ antiferro-octupole interaction ($b_6$$>$0)
effectively becomes a $\Gamma_{3g}$ antiferro-quadrupole interaction.
Thus, the $\Gamma_{3g}$ antiferro-quadrupole transition temperature
increases as $H$ is increased at a low magnetic field region.
This behavior reminds us of the experimental results for CeB$_6$,
although the order parameter in the quadrupole ordered phase of CeB$_6$
is the $\Gamma_{5g}$ quadrupole moment.
Magnetization as a function of $H$ is shown in Fig.~\ref{figure:sc_PD_M}(b).
The magnetization is isotropic as $H$$\rightarrow$0
since the $\Gamma_{4u1}$ moment is isotropic,
while anisotropy develops under a high magnetic field.

\begin{figure}[t]
  \includegraphics[width=0.9\linewidth]{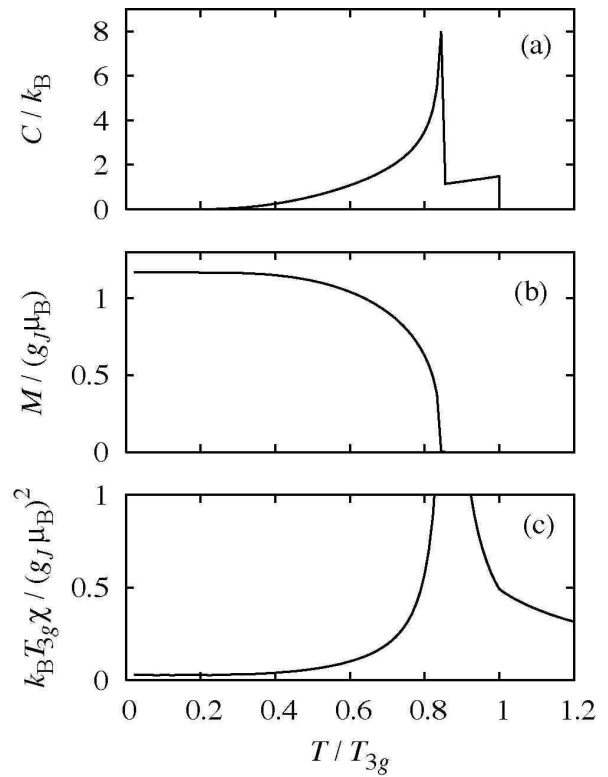}
  \caption{\label{figure:sc_physical_quant}
    Temperature dependence of physical quantities
    in the absence of magnetic field for the sc lattice.
    (a) Specific heat.
    (b) Magnetization.
    (c) Magnetic susceptibility.
  }
\end{figure}

In Figs.~\ref{figure:sc_physical_quant}(a)--(c),
we show specific heat, magnetization, and magnetic susceptibility,
respectively, as functions of temperature.
We observe two-step jump of specific heat at the quadrupole and
ferromagnetic transition temperatures, since we have applied
the mean-field theory to these second-order transitions.
Note that the magnetization starts to develop below
the ferromagnetic transition temperature.
The magnetic susceptibility exhibits a bend at $T_{3g}$,
while it diverges at the ferromagnetic transition temperature.
Under the magnetic field, this divergence turns to be a peak,
which defines the crossover to the ferromagnetic state
in the $H$-$T$ phase diagram.

Without magnetic field, the orbital ($\tau$) state is continuously
degenerate in the mean-field theory, although such continuous symmetry
is absent in this model.
As has been discussed for an $e_g$ electron model such as
perovskite manganites,\cite{Brink}
quantum fluctuations can resolve this continuous degeneracy,
but in the present model with the strong spin-orbit interaction,
magnetic field can resolve this degeneracy.
The ground states are ferromagnetic
with $\langle \mathbf{J}^{4u1} \rangle \parallel \mathbf{H}$,
where $\langle \cdots \rangle$ denotes the expectation value.
Accompanied $O^2_2$ ordering is
G-type [$\mathbf{q}=(1/2,1/2,1/2)$ in units of $2\pi/a$]
or C-type [$\mathbf{q}=(1/2,1/2,0)$]
for $\mathbf{H} \parallel [001]$,
while for $\mathbf{H} \parallel [110]$, it is C-type.
For $\mathbf{H} \parallel [111]$, there appear C-type $O^2_2$ ordering
or equivalent ones in the cubic symmetry.

\subsection{bcc lattice}
\label{sec:bcc}


The hopping integrals for the bcc lattice are given by
\begin{subequations}
\begin{align}
  t^{(a/2,a/2,a/2)} &= [\tilde{1}
    +\tilde{\tau}^y
    (+\tilde{\sigma}^x+\tilde{\sigma}^y+\tilde{\sigma}^z)/\sqrt{3}]t_2,\\
  t^{(-a/2,a/2,a/2)} &= [\tilde{1}                                     
    +\tilde{\tau}^y
    (+\tilde{\sigma}^x-\tilde{\sigma}^y-\tilde{\sigma}^z)/\sqrt{3}]t_2,\\
  t^{(a/2,-a/2,a/2)} &= [\tilde{1}                                     
    +\tilde{\tau}^y
    (-\tilde{\sigma}^x+\tilde{\sigma}^y-\tilde{\sigma}^z)/\sqrt{3}]t_2,\\
  t^{(a/2,a/2,-a/2)} &= [\tilde{1}                                     
    +\tilde{\tau}^y
    (-\tilde{\sigma}^x-\tilde{\sigma}^y+\tilde{\sigma}^z)/\sqrt{3}]t_2,
\end{align}
\end{subequations}
where $a$ is the lattice constant and $t_2$=$2(ff\sigma)/21$.


\begin{table}
  \caption{\label{table:bcc_coupling_constants}
    Coupling constants in the effective model for the bcc lattice.
    The energy unit is $(882/1327)t_2^2/(U^{\prime}-J)$.
  }
  \begin{ruledtabular}
     \begin{tabular}{ccccccccccccc}
       $a_3$ & $a_4$ & $b_5$ & $b_6$ & $b_7$ &
       $b^{(1)}_1$ & $b^{(1)}_2$ & $b^{(1)}_3$  & $b^{(1)}_4$ &
       $b^{(2)}_1$ & $b^{(2)}_2$ & $b^{(2)}_3$ & $b^{(2)}_4$ \\
          1  &  $-2$  &   9  &    2  &    2  &
              $-1$ &          2  & $-2\sqrt{3}$ &          0  &
                2  &          2  &          0  & $-2\sqrt{3}$
     \end{tabular}
  \end{ruledtabular}
\end{table}

After some algebraic calculations,
we obtain the quadrupole interaction term for the bcc lattice as
\begin{equation}
  \begin{split}
    \mathcal{H}_{1 \mathbf{q}}
    =a_3&(O_{xy, -\mathbf{q}} O_{xy, \mathbf{q}} + \text{c.p.}) c_x c_y c_z\\
    +a_4&[O_{yz, -\mathbf{q}} O_{zx, \mathbf{q}} s_x s_y c_z + \text{c.p.}].
  \end{split}
\end{equation}
The dipole and octupole interactions are given by
\begin{equation}
  \begin{split}
    \mathcal{H}_{2 \mathbf{q}}
    =b_5& T_{xyz, -\mathbf{q}} T_{xyz, \mathbf{q}} c_x c_y c_z\\
    +b_6&(T^{5u}_{z, -\mathbf{q}} T^{5u}_{z, \mathbf{q}}
      + \text{c.p.}) c_x c_y c_z \\
    +b_7&[T^{5u}_{x, -\mathbf{q}} T^{5u}_{y, \mathbf{q}} s_x s_y c_z
      + \text{c.p.}],
  \end{split}
\end{equation}
and
\begin{equation}
  \begin{split}
    \mathcal{H}_{4u n \mathbf{q}}
    =b^{(n)}_1&(J^{4u n}_{z -\mathbf{q}} J^{4u n}_{z \mathbf{q}}
      +\text{c.p.}) c_x c_y c_z \\
    +b^{(n)}_2&[J^{4u n}_{x -\mathbf{q}} J^{4u n}_{y \mathbf{q}} s_x s_y c_z
      +\text{c.p.}] \\
    +b^{(n)}_3&T_{xyz -\mathbf{q}} (J^{4u n}_{z \mathbf{q}} s_x s_y c_z
      +\text{c.p.}) \\
    +b^{(n)}_4&[T^{5u}_{z -\mathbf{q}}
      (-J^{4u n}_{x \mathbf{q}}s_z s_x c_y +J^{4u n}_{y \mathbf{q}}s_y s_z c_x)
      +\text{c.p.}],
    \label{eq:bcc_effectiveH_4u}
  \end{split}
\end{equation}
where $c_{\nu}$=$\cos(q_{\nu} a/2)$ and $s_{\nu}$=$\sin(q_{\nu} a/2)$.
The values of the coupling constants $a_i$, $b_i$, and $b^{(n)}_i$
are shown in Table~\ref{table:bcc_coupling_constants}.


In the mean-field approximation,
we find a $\Gamma_{2u}$ antiferro-octupole transition at
$T_{2u}$=$2b_5/k_{\text{B}}$ with $\mathbf{q}$=$(1,0,0)$,
and a $\Gamma_{4u1}$ ferromagnetic transition at a lower temperature.
The ground state has the $\Gamma_{5g}$ antiferro-quadrupole moment
with the same ordering wave-vector as the $\Gamma_{2u}$ moment.
The ground state energy is $-a_3-b_5+b^{(1)}_1$ per site.

\begin{figure}
  \includegraphics[width=0.9\linewidth]{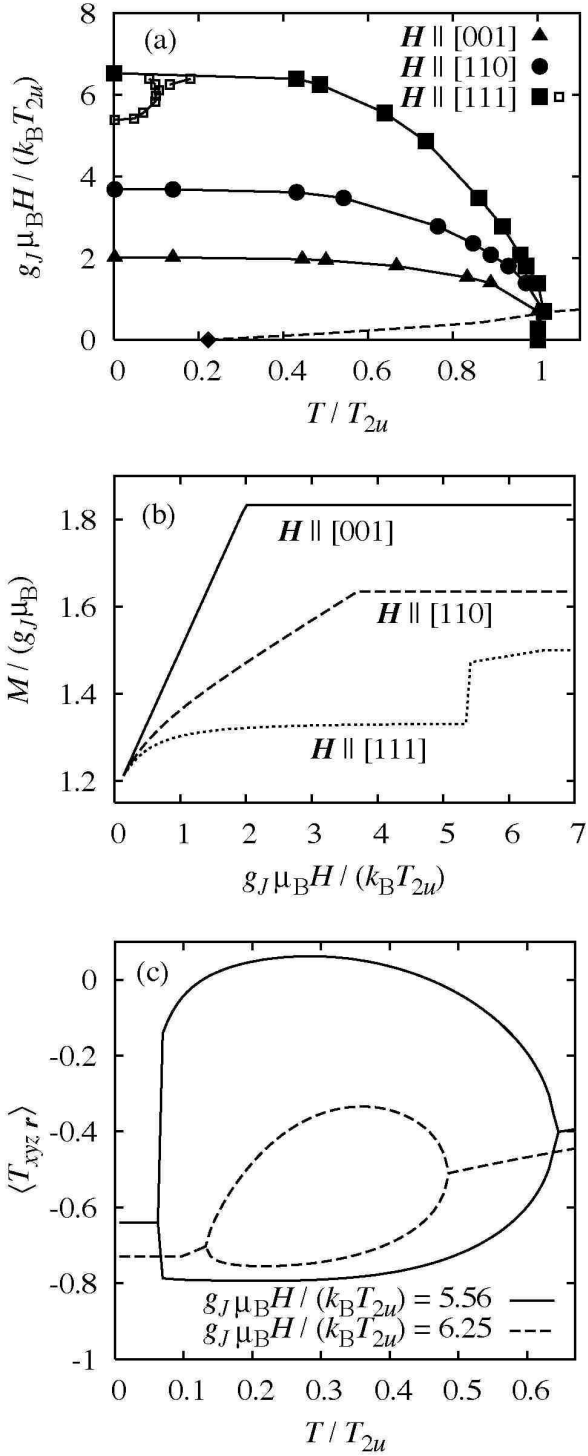}
  \caption{\label{figure:bcc_PD_M}
  (a) $H$-$T$ phase diagram for the bcc lattice
  for three magnetic field directions.
  Solid symbols denote the $\Gamma_{2u}$ octupole transition.
  The diamond represents the ferromagnetic transition point.
  The dashed curve represents the crossover to the ferromagnetic state.
  As for the definition of the crossover, see the main text.
  Open rectangles for $\mathbf{H} \parallel [111]$ denote
  transitions to uniform phases, as shown in (c).
  (b) Magnetization as a function of magnetic field
  for the bcc lattice.
  (c) Expectation value of $\Gamma_{2u}$ octupole moment
  $\langle T_{xyz \mathbf{r}} \rangle$ at each of sublattice sites
  $\mathbf{r}=(0,0,0)$ and $(a/2,a/2,a/2)$
  under high magnetic fields along [111] for the bcc lattice.
}
\end{figure}

In Fig.~\ref{figure:bcc_PD_M}(a), we show an $H$-$T$ phase diagram.
Again the ferromagnetic transition becomes a crossover under
the finite magnetic field.
The crossover curve determined by the peak in the magnetic susceptibility
is found to be almost isotropic in the region shown here.
Then, we show only the curve for $\mathbf{H} \parallel [001]$.
In the region for high $H$ and low $T$, we find two uniform phases.
One is a phase with uniform $\langle T_{xyz \mathbf{r}} \rangle$
depending on temperature and another is a phase with
uniform $\langle T_{xyz \mathbf{r}} \rangle$ which does not depend
on temperature, as shown in Figs.~\ref{figure:bcc_PD_M}(a) and (c).

In Fig.~\ref{figure:bcc_PD_M}(b),
we show magnetization as a function of $H$.
We note that the magnetization is isotropic as $H$$\rightarrow$0
as in the sc lattice, since the order parameter of the ferromagnetic
transition is the $\Gamma_{4u1}$ moment.
Note also that the jump in the magnetization
at $(g_J \mu_{\text{B}} H)/(k_{\text{B}}T_{2u})$=5.4
for $\mathbf{H} \parallel [111]$
indicates the transition to the uniform state.

\begin{figure}
  \includegraphics[width=0.9\linewidth]{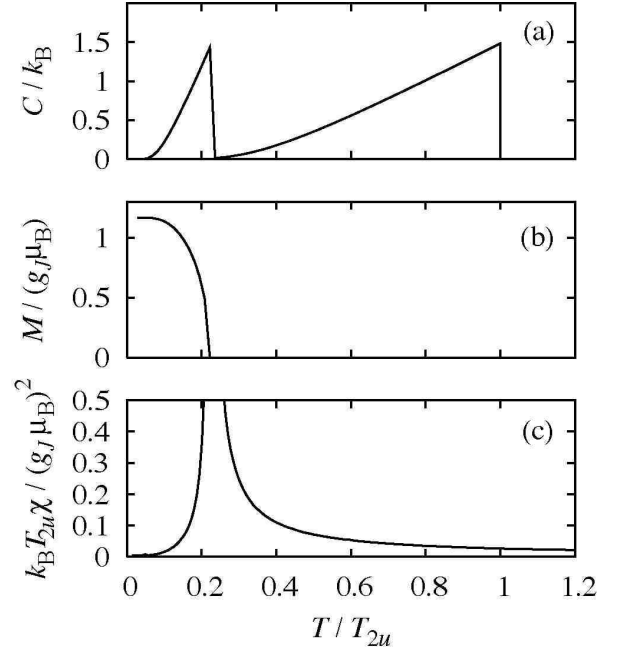}
  \caption{\label{figure:bcc_physical_quant}
    Temperature dependence of physical quantities
    in the absence of magnetic field for the bcc lattice.
    (a) Specific heat.
    (b) Magnetization.
    (c) Magnetic susceptibility.
  }
\end{figure}

Figures~\ref{figure:bcc_physical_quant}(a)--(c) show
specific heat, magnetization and magnetic susceptibility
as functions of temperature, respectively.
We observe two jumps in the specific heat at the octupole and
ferromagnetic transition temperatures.
The magnetization begins to develop below the ferromagnetic
transition temperature.
The magnetic susceptibility has a bend at $T_{2u}$
and diverges at the ferromagnetic transition temperature.
Note that the anomaly in the magnetic susceptibility at $T_{2u}$
is very weak.
In the pure magnetic $\Gamma_{2u}$ octupole ordered state,
there remains degeneracy, while in ordinary magnetic states,
degeneracy is fully resolved.
Thus, the nature of the $\Gamma_{2u}$ octupole phase is similar
to that of the quadrupole ordered phases.
For instance, the anomaly in the magnetic susceptibility is weak
at the transition temperature,
there is no ordered magnetic dipole moment,
and another phase transition occurs at a lower temperature.


The ground state is continuously degenerate,
since the $\Gamma_{4u1}$ and $\Gamma_{5g}$ moments are isotropic
in this model within the $\mathbf{q}=(1,0,0)$ structure.
We note that this degeneracy is due to the symmetry of the model
in contrast to the sc lattice.
By applying a magnetic field, the ground states are uniquely determined.
The ground states are ferromagnetic phases
$\langle \mathbf{J}^{4u1} \rangle \parallel \mathbf{H}$
with antiferro $O_{xy}$ ordering for $\mathbf{H} \parallel [001]$,
with antiferro $O_{yz}+O_{zx}$ ordering for $\mathbf{H} \parallel [110]$,
and with antiferro $O_{yz}+O_{zx}+O_{xy}$ ordering for
$\mathbf{H} \parallel [111]$.

As mentioned in Sec.~I,
quite recently, a possibility of octupole ordering in
filled skutterudite compound SmRu$_4$P$_{12}$ has been suggested
experimentally.\cite{Yoshizawa,Hachitani}
In the filled skutterudite structure, rare-earth ion surrounded
by pnictogens form the bcc lattice.
Moreover, the $\Gamma_8$ CEF ground state has been reported
in the Sm-based filled skutterudite.~\cite{Matsuhira}
Thus, we expect to apply the present model to
Sm-based filled skutterudites.
When we compare our result on the bcc lattice
with the experimental suggestion,
octupole ordering actually occurs in our model for the bcc lattice,
but $\Gamma_{2u}$ octupole ordered state does not seem to explain
the experimental results.
This discrepancy is due to the suppression of $\Gamma_7$ orbital,
since in the filled skutterudites, conduction electron has
$a_u$ symmetry, which hybridizes with $\Gamma_7$ electron.
In addition, the level splitting between $\Gamma_7$ and $\Gamma_8$
is considered to be rather small in filled skutterudites.
Thus, for filled-skutterudite materials, we should consider
the $j$=5/2 sextet model in the bcc lattice with itinerant
$\Gamma_7$ and localized $\Gamma_8$ orbitals.
We postpone the analysis of such a model in future.

\subsection{fcc lattice}
\label{sec:fcc}


The hopping integrals for the fcc lattice are given by
\begin{subequations}
\begin{align}
t^{(0,a/2,a/2)}&=
[\tilde{1}+(\tilde{\eta}^+ -4\sqrt{3} \tilde{\tau}^y
\tilde{\sigma}^x)/7]t_3, \\
t^{(a/2,0,a/2)}&=                                                  
[\tilde{1}+(\tilde{\eta}^- -4\sqrt{3} \tilde{\tau}^y
\tilde{\sigma}^y)/7]t_3, \\
t^{(a/2,a/2,0)}&=                                                  
[\tilde{1}+(\tilde{\tau}^z -4\sqrt{3} \tilde{\tau}^y
\tilde{\sigma}^z)/7]t_3, \\
t^{(0,a/2,-a/2)}&=                                                 
[\tilde{1}+(\tilde{\eta}^+ +4\sqrt{3} \tilde{\tau}^y
\tilde{\sigma}^x)/7]t_3, \\
t^{(-a/2,0,a/2)}&=                                                 
[\tilde{1}+(\tilde{\eta}^- +4\sqrt{3} \tilde{\tau}^y
\tilde{\sigma}^y)/7]t_3, \\
t^{(a/2,-a/2,0)}&=                                                 
[\tilde{1}+(\tilde{\tau}^z +4\sqrt{3} \tilde{\tau}^y
\tilde{\sigma}^z)/7]t_3,
\end{align}
\label{eqs:hopping:fcc}
\end{subequations}
where $a$ is the lattice constant and $t_3$=$(ff\sigma)/8$.


Each multipole interaction term in the effective Hamiltonian for the fcc
lattice is given by
\begin{equation}
  \begin{split}
    \mathcal{H}_{1 \mathbf{q}}
    =a_1&(O^0_{2, -\mathbf{q}} O^0_{2, \mathbf{q}} c_x c_y + \text{c.p.})\\
    +a_3&(O^0_{2, -\mathbf{q}} O_{xy, \mathbf{q}} s_x s_y + \text{c.p.})\\
    +a_4&(O_{xy, -\mathbf{q}} O_{xy, \mathbf{q}} c_x c_y + \text{c.p.}),
  \end{split}
  \label{eq:fcc_effectiveH_even}
\end{equation}
\begin{equation}
  \begin{split}
    \mathcal{H}_{2 \mathbf{q}}
    &=b_8[T^{5u}_{z, -\mathbf{q}} T^{5u}_{z, \mathbf{q}} (c_y c_z + c_z c_x)
      + \text{c.p.}] \\
    &+b_9[T^{5u}_{x, -\mathbf{q}} T^{5u}_{y, \mathbf{q}} s_x s_y
      + \text{c.p.}] \\
    &+b_{10} T_{xyz, -\mathbf{q}} T_{xyz, \mathbf{q}} (c_x c_y + \text{c.p.}),
  \end{split}
\end{equation}
and
\begin{equation}
  \begin{split}
    \mathcal{H}_{4u n \mathbf{q}}
    =b^{(n)}_1&[J^{4u n}_{z -\mathbf{q}} J^{4u n}_{z \mathbf{q}} c_x c_y
      +\text{c.p.}] \\
    +b^{(n)}_2&[J^{4u n}_{z -\mathbf{q}} J^{4u n}_{z \mathbf{q}}
      (c_y c_z +c_z c_x)+\text{c.p.}] \\
    +b^{(n)}_3&[J^{4u n}_{x -\mathbf{q}} J^{4u n}_{y \mathbf{q}} s_x s_y
      +\text{c.p.}] \\
    +b^{(n)}_4&[T_{xyz -\mathbf{q}} (J^{4u n}_{z \mathbf{q}} s_x s_y
      +\text{c.p.})] \\
    +b^{(n)}_5&[T^{5u}_{z -\mathbf{q}} J^{4u n}_{z \mathbf{q}} c_z(c_x-c_y)
      +\text{c.p.})] \\
    +b^{(n)}_6&[T^{5u}_{z -\mathbf{q}}
      (-J^{4u n}_{x \mathbf{q}}s_z s_x+J^{4u n}_{y \mathbf{q}}s_y s_z)
      +\text{c.p.}].
    \label{eq:fcc_effectiveH_4u}
  \end{split}
\end{equation}
The values of the coupling constants $a_i$, $b_i$ and $b^{(n)}_i$
are shown in Table~\ref{table:fcc_coupling_constants}.

\begin{table}
  \caption{\label{table:fcc_coupling_constants}
    Coupling constants in the effective model for the fcc lattice.
    The energy unit is $(1/196)t_3^2/(U^{\prime}-J)$.
  }
  \begin{ruledtabular}
    \begin{tabular}{ccccccccc}
      $a_1$       & $a_3$        & $a_4$ &
      $b_8$       & $b_9$        & $b_{10}$ &
      $b^{(1)}_1$ & $b^{(1)}_2$  & $b^{(1)}_3$ \\
      \vspace{3mm}
      12          & $64\sqrt{3}$ & 192 &
      195         & $-336$       & 576 &
      $-196$      & $-4$         & 0           \\
      $b^{(1)}_4$   & $b^{(1)}_5$ & $b^{(1)}_6$ &
      $b^{(2)}_1$   & $b^{(2)}_2$ & $b^{(2)}_3$ &
      $b^{(2)}_4$   & $b^{(2)}_5$ & $b^{(2)}_6$ \\
      $224\sqrt{3}$ & 0           & 0           &
      4             & 193         & $-336$      &
      $64\sqrt{3}$  & $2\sqrt{3}$ & $112\sqrt{3}$
    \end{tabular}
  \end{ruledtabular}
\end{table}

\begin{figure}
  \includegraphics[width=1\linewidth]{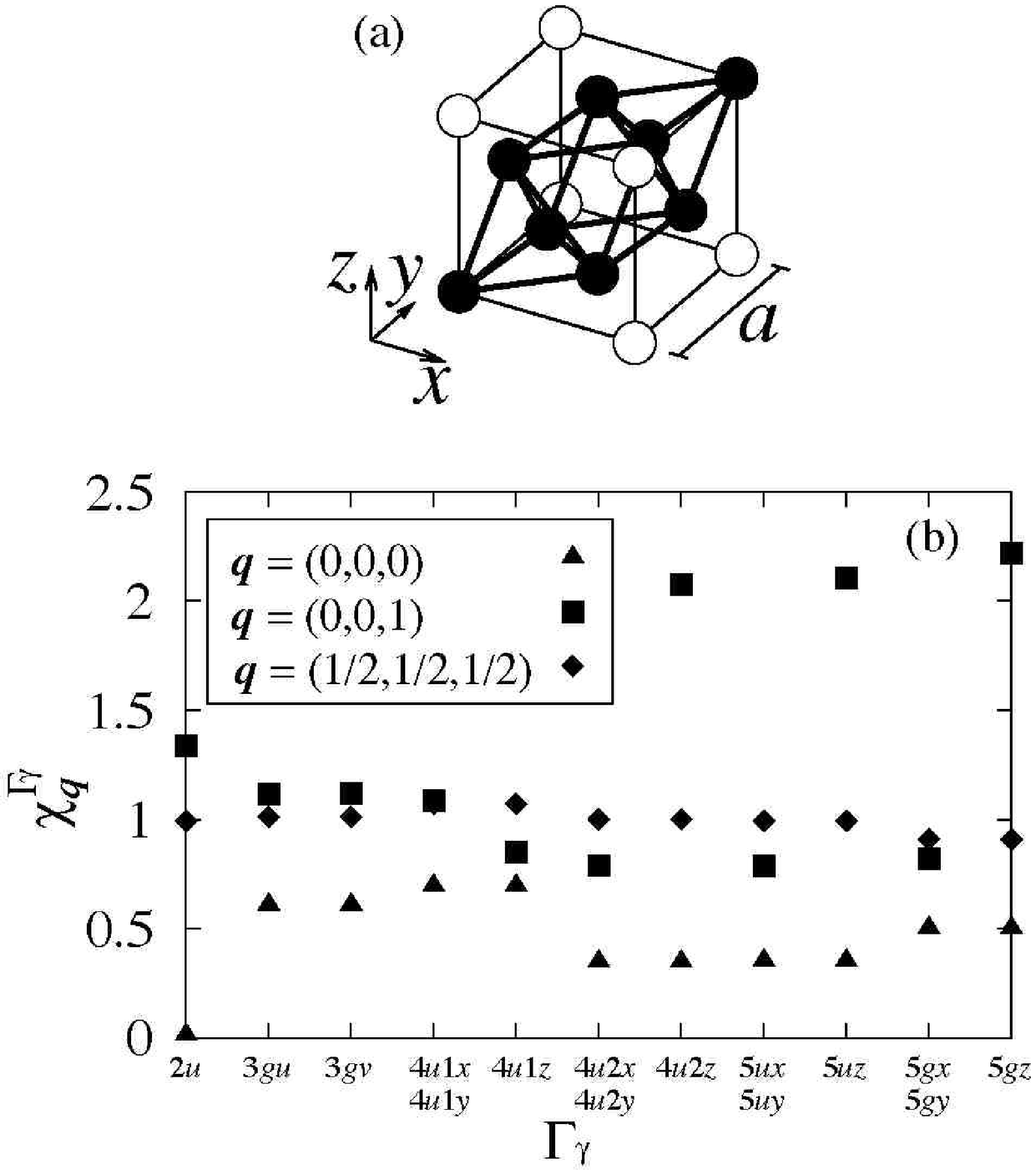}
  \caption{\label{figure:correlation_function}
    (a) 8-site cluster (solid spheres) on the fcc lattice
    taken in the calculation.
    (b) Correlation functions for the 8-site cluster.
    The unit of the wave vectors is  $2\pi/a$.
  }
\end{figure}

As already mentioned in Ref.~\onlinecite{Kubo:NpO2}, it is necessary
to analyze the effective model carefully for the fcc lattice,
since the model includes geometrical frustration.
It is risky to apply directly the mean-field approximation
to the effective model.
First we evaluate the correlation function in the ground state using
an unbiased method such as exact diagonalization on the $N$-site lattice.
Here we set $N$=8, as shown in Fig.~\ref{figure:correlation_function}(a).
The correlation function of the multipole operators is given by
\begin{equation}
  \chi^{\Gamma_{\gamma}}_{\mathbf{q}}=(1/N)
  \sum_{\mathbf{r},\mathbf{r}^{\prime}}
  e^{i \mathbf{q} \cdot (\mathbf{r}-\mathbf{r}^{\prime})}
  \langle X^{\Gamma_{\gamma}}_{\mathbf{r}}
  X^{\Gamma_{\gamma}}_{\mathbf{r}^{\prime}} \rangle,
\end{equation}
where $\langle \cdots \rangle$ denotes the expectation value
using the ground-state wave-function.

In Fig.~\ref{figure:correlation_function} (b),
we show results for the correlation functions.
The interaction between $\Gamma_{2u}$ moments ($b_{10}$) is large,
but the correlation function of the $\Gamma_{2u}$ moment is not enhanced,
indicating that the frustration effect is significant
for an Ising-like moment such as $\Gamma_{2u}$.
We find large values of correlation functions for $J^{4u2}_{z}$,
$T^{5u}_z$, and $O_{xy}$ moments at $\mathbf{q}$=$(0,0,1)$.
However, there is no term in the effective model
which stabilizes $O_{xy}$ quadrupole order at $\mathbf{q}$=$(0,0,1)$.
We note that either of $\Gamma_{4u2}$ and $\Gamma_{5u}$ ordered states
can accompany $\Gamma_{5g}$ quadrupole moments.
Thus, the enhancement of $O_{xy}$ correlation function indicates
an induced quadrupole moment in $\Gamma_{4u2}$ or $\Gamma_{5u}$ moment
ordered states.
Namely, the relevant interactions are $b^{(2)}_2$ and $b_8$,
which stabilize the $J^{4u2}_z$ and $T^{5u}_z$ order, respectively,
at $\mathbf{q}$=$(0,0,1)$.

Next we study the ordered state by applying mean-field theory
to the simplified model including only $b^{(2)}_2$ and $b_8$.
Since the coupling constant $b_8$ is slightly larger than $b^{(2)}_2$,
$\Gamma_{5u}$ ordered state should has lower energy than
$\Gamma_{4u2}$ ordered state.
The interaction $b_8$ stabilizes longitudinal ordering of
the $\Gamma_{5u}$ moments, i.e.,
$\langle \mathbf{T}^{5u}_{\mathbf{r}} \rangle
\parallel \mathbf{q}$.

However, we cannot conclude that
the ground state is the single-$\mathbf{q}$ state
$(\langle T^{5u}_{x \mathbf{r}} \rangle,
\langle T^{5u}_{y \mathbf{r}} \rangle,
\langle T^{5u}_{z \mathbf{r}} \rangle) \propto
(0,0,\exp[i 2 \pi z/a])$,
since there is a possibility of multi-$\mathbf{q}$ structures.
For isotropic moments, single-$\mathbf{q}$ and multi-$\mathbf{q}$
structures have the same energy, and thus, anisotropy in the moment
is important to determine the stable structure.
Indeed, the $\Gamma_{5u}$ moment has an easy axis along [111]
in the $\Gamma_8$ subspace.~\cite{Kubo,Kiss}
In this case, we find that a triple-$\mathbf{q}$ state is most stable
among the single-$\mathbf{q}$ and multi-$\mathbf{q}$ states,
since it gains interaction energy in all the directions.

\begin{figure}
  \includegraphics[width=1\linewidth]{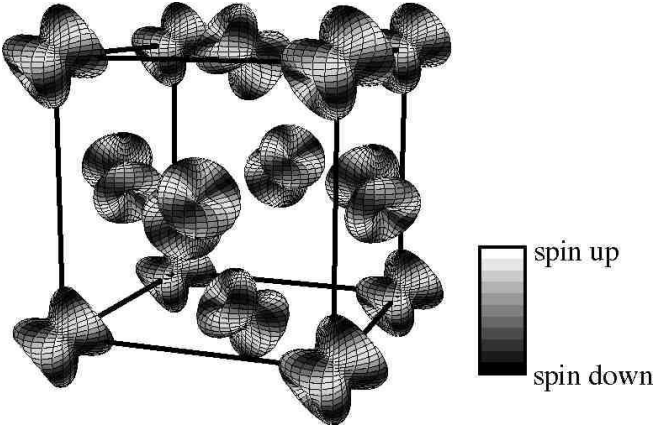}
  \caption{\label{figure:fcc_triple_q}
  The triple-$\mathbf{q}$ $\Gamma_{5u}$ octupole state.
  The surface is defined by
  $r$=$[\sum_{\sigma}|\psi(\theta,\phi,\sigma)|^2]^{1/2}$
  in the polar coordinates,
  when the $5f$ wave-function is represented by
  $\Psi(r, \theta, \phi, \sigma)$=$R(r)\psi(\theta,\phi,\sigma)$,
  where $\sigma$ denotes real spin.
  White-shift of the surface indicates the increase of the weight of
  up-spin state $|\psi(\theta,\phi,\uparrow)|^2/r^2$.
}
\end{figure}

In fact, the mean-field ground-state of the simplified model is
the longitudinal triple-$\mathbf{q}$ $\Gamma_{5u}$ octupole state
with four sublattices, i.e.,
\begin{subequations}
  \begin{align}
    \langle T^{5u}_{x \mathbf{r}} \rangle &\propto \exp[i 2 \pi x/a], \\
    \langle T^{5u}_{y \mathbf{r}} \rangle &\propto \exp[i 2 \pi y/a], \\
    \langle T^{5u}_{z \mathbf{r}} \rangle &\propto \exp[i 2 \pi z/a].
  \end{align}
\end{subequations}
This state accompanies the triple-$\mathbf{q}$ quadrupole moment~\cite{Paixao}
\begin{subequations}
\begin{align}
  \langle O_{yz \mathbf{r}} \rangle
  &\propto \langle T^{5u}_{x \mathbf{r}} \rangle, \\
  \langle O_{zx \mathbf{r}} \rangle
  &\propto \langle T^{5u}_{y \mathbf{r}} \rangle, \\
  \langle O_{xy \mathbf{r}} \rangle
  &\propto \langle T^{5u}_{z \mathbf{r}} \rangle.
\end{align}
\end{subequations}
In Fig.~\ref{figure:fcc_triple_q},
we show symmetry of the charge distribution with spin density
in the triple-$\mathbf{q}$ $\Gamma_{5u}$ octupole state.
Note that this triple-$\mathbf{q}$ structure does not have frustration
even in the fcc lattice.
The ground state energy is $-4b_8$ per site,
and the transition temperature is given by $k_{\text{B}} T_{5u}$=$4b_8$.
We also note that this triple-$\mathbf{q}$ $\Gamma_{5u}$ octupole state
has been proposed for NpO$_2$ phenomenologically.~\cite{Paixao}

\begin{figure}
  \includegraphics[width=0.9\linewidth]{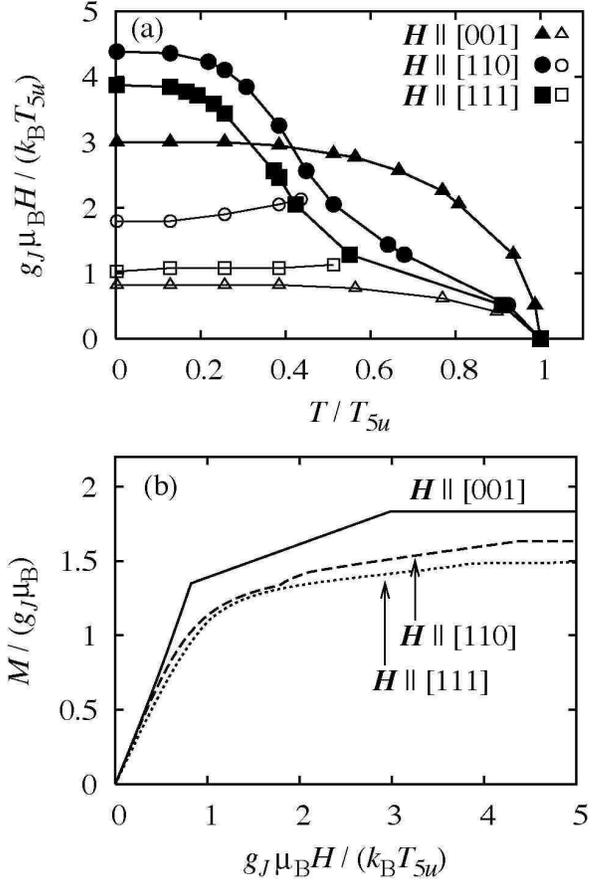}
  \caption{\label{figure:fcc_PD_M}
   Phase diagram and magnetization for
   the simplified model on the fcc lattice.
   (a) $H$-$T$ phase diagram.
   Solid symbols denote the $\Gamma_{5u}$ octupole transition,
   while open symbols denote transitions between
   $\Gamma_{5u}$ octupole ordered states
   with different sublattice structures (see Fig.~\ref{figure:fcc_T5u}).
   (b) Magnetic field dependence of the magnetization.
  }
\end{figure}

\begin{figure}
  \includegraphics[width=1\linewidth]{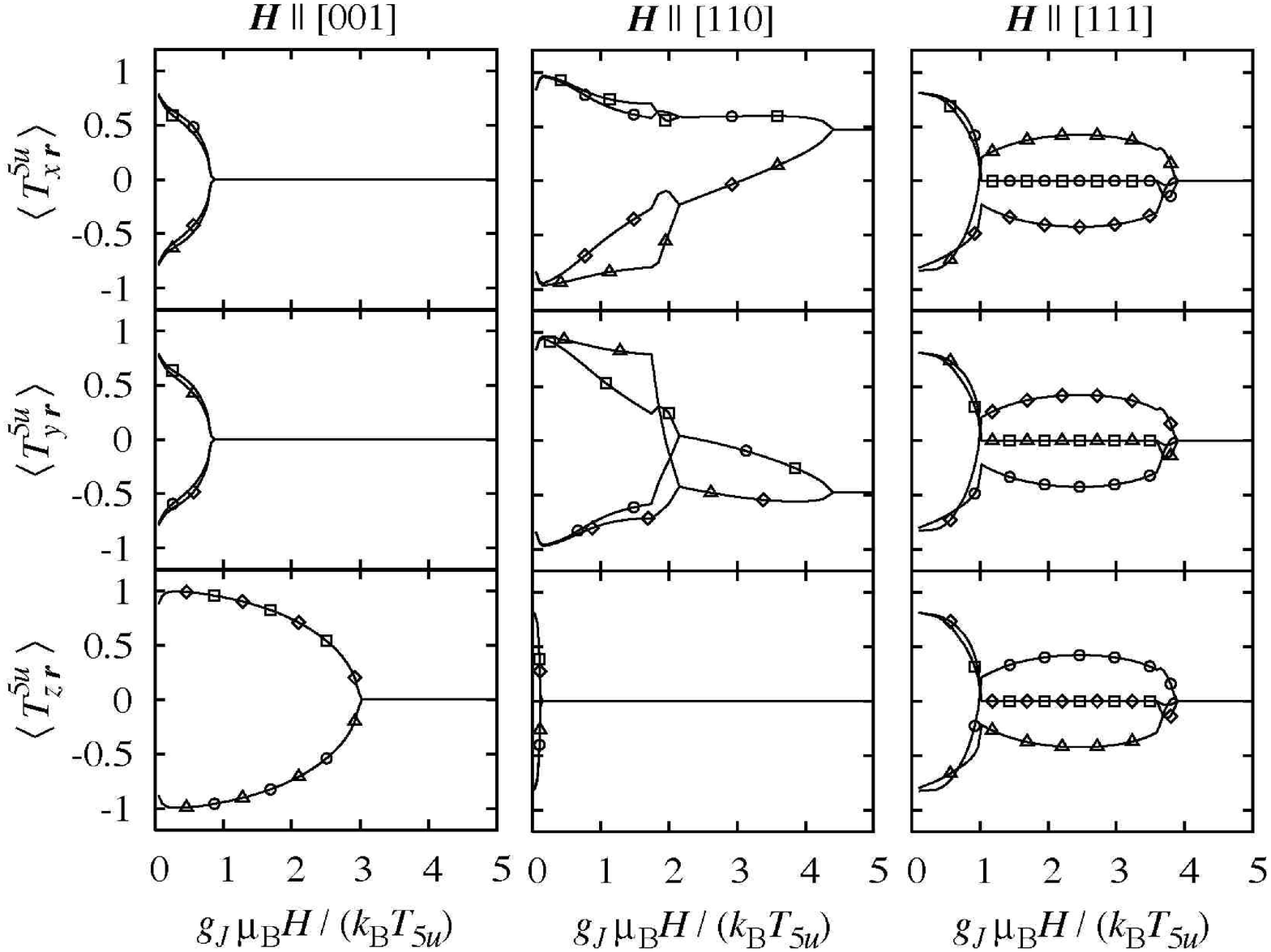}
  \caption{\label{figure:fcc_T5u}
    Magnetic field dependence of $\Gamma_{5u}$ octupole moments
    at $T$=0 at each of four-sublattice sites:
    $\mathbf{r}$=$(0,0,0)$ (open rectangles),
    $\mathbf{r}$=$(0,a/2,a/2)$ (open circles),
    $\mathbf{r}$=$(a/2,0,a/2)$ (open triangles), and
    $\mathbf{r}$=$(a/2,a/2,0)$ (open diamonds) for the fcc lattice.
  }
\end{figure}

Let us now evaluate physical quantities in the mean-field theory.
Figures~\ref{figure:fcc_PD_M}(a) and (b) show an $H$-$T$ phase diagram and
the magnetic field dependence of the magnetization at $T$=0, respectively.
Note that the magnetization is isotropic as $H$$\rightarrow$0
due to the cubic symmetry.
The bend for $\mathbf{H} \parallel [001]$
and the dip for $\mathbf{H} \parallel [110]$
in magnetization indicate transitions to the two-sublattice structures.
There is anomaly in magnetization also for $\mathbf{H} \parallel [111]$
at the transition to the different sublattice structure,
but it is very weak.

Under a high magnetic field, sublattice structures change,
as shown in Fig.~\ref{figure:fcc_T5u}:
For $\mathbf{H} \parallel [001]$,
we obtain a two-sublattice structure with
\begin{subequations}
  \begin{align}
    \langle T^{5u}_{x \mathbf{r}} \rangle &= 0, \\
    \langle T^{5u}_{y \mathbf{r}} \rangle &= 0, \\
    \langle T^{5u}_{z \mathbf{r}} \rangle &\propto \exp[i2\pi z/a].
  \end{align}
\end{subequations}
For $\mathbf{H} \parallel [110]$, there appears
a two-sublattice structure with
\begin{subequations}
  \begin{align}
    \langle T^{5u}_{x \mathbf{r}} \rangle &\ne 0, \\
    \langle T^{5u}_{y \mathbf{r}} \rangle &\ne 0, \\
    \langle T^{5u}_{z \mathbf{r}} \rangle &= 0.
  \end{align}
\end{subequations}
Finally, for $\mathbf{H} \parallel [111]$, we observe
\begin{subequations}
  \begin{align}
    \langle T^{5u}_{x \mathbf{r}} \rangle
    &\propto \sin[2\pi (y-z)/a], \\
    \langle T^{5u}_{y \mathbf{r}} \rangle
    &\propto \sin[2\pi (z-x)/a], \\
    \langle T^{5u}_{z \mathbf{r}} \rangle
    &\propto \sin[2\pi (x-y)/a].
  \end{align}
\end{subequations}
Note also that the triple-$\mathbf{q}$ state
is fragile under $\mathbf{H} \parallel [110]$:
$\langle T^{5u}_{z \mathbf{r}} \rangle$=0 with
a four-sublattice structure for
$g_J \mu_{\text{B}} H/(k_{\text{B}} T_{5u}) \gtrsim 0.11$ at $T$=0
[this phase boundary is not shown in Fig.~\ref{figure:fcc_PD_M}(a)].

\begin{figure}
  \includegraphics[width=0.9\linewidth]{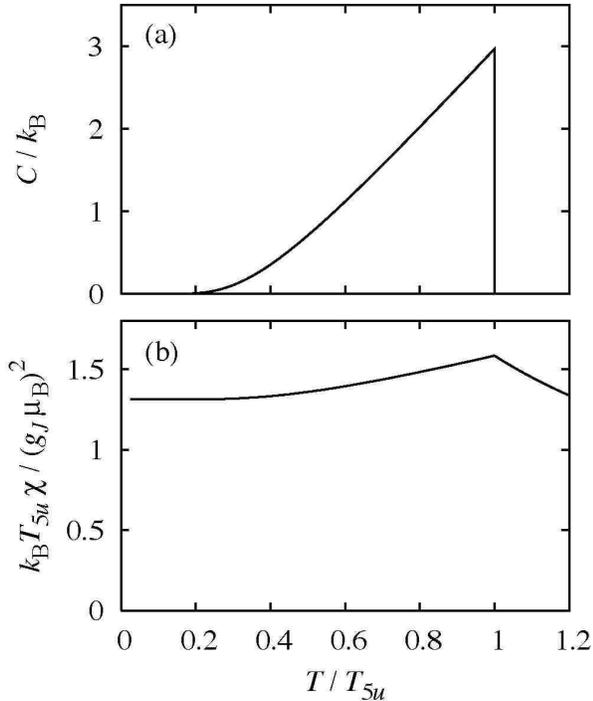}
  \caption{\label{figure:fcc_physical_quant}
    Temperature dependence of physical quantities 
    in the absence of a magnetic field for the fcc lattice.
    (a) Specific heat.
    (b) Magnetic susceptibility.
  }
\end{figure}

Figures~\ref{figure:fcc_physical_quant}(a)
and \ref{figure:fcc_physical_quant}(b) show
the temperature dependence of the specific heat
and magnetic susceptibility, respectively.
At $T$=$T_{5u}$, there appear the specific heat jump
and a cusp in the magnetic susceptibility.
In contrast to the sc and bcc lattices,
there occurs single phase transition at zero magnetic field
in the case of the fcc lattice.
Note also that the cusp structure in the magnetic susceptibility
is rather strong compared with experimental results.\cite{Kubo:NpO2,Ross}
Such a quantitative disagreement with experiments is considered to
originate from the suppression of $\Gamma_7$ orbital in our model.
The analysis of the $j$=5/2 sextet model on the fcc lattice
is one of future problems.

\section{Discussion and summary}
\label{sec:summary}

We have constructed $\Gamma_8$ models with hopping integrals
through $(ff\sigma)$ bonding based on the $j$-$j$ coupling scheme.
In order to study multipole ordering, we have derived an effective model
by using the second-order perturbation theory
with respect to $f$-$f$ hopping.
By applying mean-field theory, we find different multipole ordered states
depending on the lattice structure.
For the sc lattice, a $\Gamma_{3g}$ antiferro-quadrupole transition
occurs at a finite temperature.
As lowering temperature further, we find a ferromagnetic transition.
For the bcc lattice, a $\Gamma_{2u}$ antiferro-octupole ordering
occurs first, and a ferromagnetic transition follows it.
Finally, for the fcc lattice, with careful analysis, we conclude
the appearance of the single phase transition to
the triple-$\mathbf{q}$ $\Gamma_{5u}$ octupole ordering.

In this paper, we have not taken into account
the effect of conduction electron.
One may complain about this point, since it is believed
that the hybridization of $f$ electrons with conduction electron band
is important to understand the magnetism of $f$-electron systems.
In fact, in the traditional prescription, first we derive
the Coqblin-Schrieffer model from the periodic Anderson model
by evaluating the $c$-$f$ exchange interaction $J_{\rm cf}$
within the second-order perturbation in terms of the hybridization
between $f$- and conduction electrons.\cite{Coqblin}
Then, we derive the RKKY interactions again using the second-order
perturbation theory with respect to $J_{\rm cf}$.

In general, the RKKY interactions are orbital dependent
and interpreted as multipole interactions.
Such orbital dependence originates from that of the hybridization.
Note that the hybridization should occur only between $f$- and
conduction band with the same symmetry.
Here we emphasize that the symmetry of $f$-electron state is
correctly included in our calculations.
Thus, the structure in the multipole interactions will not be
changed so much, even if we consider the effect of hybridization
with conduction band, as long as we consider correctly
the symmetry of $f$ electron states.

Let us show an example to support our belief.
Concerning the octupole ordering in NpO$_2$,
we have extended the present theory by further including
the effect of $p$ electrons of oxygen anions.\cite{Kubo:fp}
Namely, we have constructed the so-called $f$-$p$ model,
given in the form of
\begin{equation}
\mathcal{H} =\mathcal{H}_{\rm f}
  +\mathcal{H}_{\rm p}
  +\mathcal{H}_{\rm hyb},
\end{equation}
where $\mathcal{H}_{\rm f}$ and $\mathcal{H}_{\rm p}$ denote
the local $f$- and $p$-electron terms, respectively, and
$\mathcal{H}_{\rm hyb}$ is the hybridization between $p$- and
$f$-electrons through $(pf\sigma)$ and $(pf\pi)$.
Then, it has been found that the structure in the multipole interactions
of the effective model derived from the $f$-$p$ model is qualitatively
the same as those obtained in the $\Gamma_8$ model on the fcc lattice.
In fact, we have found a finite parameter region
of $\Gamma_{5u}$ antiferro-octupole phase.
Namely, the $f$-$p$ model on the fcc lattice has a
tendency toward $\Gamma_{5u}$ antiferro-octupole ordering,
which has been already captured in the simple $(ff\sigma)$ model.
This result suggests that the structure in multipole interactions 
is determined mainly by the symmetry of $f$-electron state.
Most of the effect of hybridization can be included by changing
effectively $(ff\sigma)$ in the multipole interactions
shown in the present paper.

However, if the itinerant nature of $f$ electrons is increased
due to the large hybridization and metallicity of the ground state
becomes significant, the present approximation inevitably loses
the validity and the effect of the conduction band should
be important.
In such a case, it is necessary to develop a theory on the
basis of the orbital-degenerate periodic Anderson model
in order to include the multipole fluctuations.
It is one of future tasks.

\section*{Acknowledgments}

We thank M. Yoshizawa and H. Fukazawa for sending us preprints
on SmRu$_4$P$_{12}$ prior to publication.
We also thank H. Harima, S. Kambe, N. Metoki, Y. Tokunaga, K. Ueda,
R. E. Walstedt, and H. Yasuoka for useful discussions.
One of the authors (K. K.) is grateful to H. Onishi for useful comments
on numerical diagonalization.
K. K. is supported by the REIMEI Research Resources of Japan Atomic Energy
Research Institute.
Another author (T. H.) is supported by a Grants-in-Aid
for Scientific Research in Priority Area ``Skutterudites''
under the contract No.~16037217 from the Ministry of
Education, Culture, Sports, Science, and Technology of Japan.
T. H. is also supported by a Grant-in-Aid for
Scientific Research (C)(2) under the contract No.~50211496
from Japan Society for the Promotion of Science.


\end{document}